\documentclass[12pt,english]{article}
\pdfoutput=1
\usepackage{lmodern}

\usepackage[T1]{fontenc}
\usepackage[latin9]{inputenc}
\usepackage{geometry}
\usepackage{color}
\usepackage{babel}
\usepackage{cite}
\usepackage{mathtools}
\usepackage{amsmath}
\usepackage{amssymb}
\usepackage{graphicx}
\usepackage{esint}
\usepackage{bbm}
\usepackage[unicode=true,pdfusetitle,
 bookmarks=true,bookmarksnumbered=false,bookmarksopen=false,
 breaklinks=false,pdfborder={0 0 1},backref=false,colorlinks=true]
 {hyperref}
\hypersetup{
 citecolor=black,linkcolor=black,urlcolor=black}
\makeatletter

 \@ifundefined{textcolor}{}
 {
   \definecolor{BLACK}{gray}{0}
   \definecolor{WHITE}{gray}{1}
   \definecolor{RED}{rgb}{1,0,0}
   \definecolor{GREEN}{rgb}{0,1,0}
   \definecolor{BLUE}{rgb}{0,0,1}
   \definecolor{CYAN}{cmyk}{1,0,0,0}
   \definecolor{MAGENTA}{cmyk}{0,1,0,0}
   \definecolor{YELLOW}{cmyk}{0,0,1,0}
 }
\usepackage{babel}
\makeatletter
\def\simgt{\mathrel{\lower2.5pt\vbox{\lineskip=0pt\baselineskip=0pt
           \hbox{$>$}\hbox{$\sim$}}}}
\def\simlt{\mathrel{\lower2.5pt\vbox{\lineskip=0pt\baselineskip=0pt
           \hbox{$<$}\hbox{$\sim$}}}}
\makeatother

\newcommand{\be}{\begin{equation}}
\newcommand{\ee}{\end{equation}}
\newcommand{\bea}{\begin{eqnarray}}
\newcommand{\eea}{\end{eqnarray}}

\newcommand{\Ref}[1]{Ref.~\cite{#1}}

\newcommand{\Eq}[1]{Eq.~\eqref{#1}}
\newcommand{\Eqs}[2]{Eqs.~\eqref{#1} and \eqref{#2}}
\newcommand{\Sec}[1]{Sec.~\ref{#1}}
\newcommand{\Secs}[2]{Secs.~\ref{#1} and \ref{#2}}

\newcommand{\LGF}{{\cal L}_{\rm GF}}

\allowdisplaybreaks

\begin{document}
\interfootnotelinepenalty=10000
\baselineskip=18pt

\hfill CALT-TH-2016-037
\hfill

\vspace{2cm}
\thispagestyle{empty}
\begin{center}
{\Large \bf
Twofold Symmetries of the Pure Gravity Action
}\\
\bigskip\vspace{1cm}{
{\large Clifford Cheung and Grant N. Remmen}
} \\[7mm]
 {\it Walter Burke Institute for Theoretical Physics \\[-1mm]
    California Institute of Technology, Pasadena, CA 91125} \let\thefootnote\relax\footnote{e-mail:
\url{clifford.cheung@caltech.edu}, \url{gremmen@theory.caltech.edu}} \\
 \end{center}
\bigskip
\centerline{\large\bf Abstract}

\begin{quote} \small

We recast the action of pure gravity into a form that is invariant under a twofold Lorentz symmetry. 
To derive this representation, we construct a general parameterization of all theories equivalent to the Einstein-Hilbert action up to a local field redefinition and gauge fixing.  We then exploit this freedom to eliminate all interactions except those exhibiting two sets of independently contracted Lorentz indices.  The resulting action is local, remarkably simple, and naturally expressed in a field basis analogous to the exponential parameterization of the nonlinear sigma model.  The space of twofold Lorentz invariant field redefinitions then generates an infinite class of equivalent representations.  By construction, all off-shell Feynman diagrams are twofold Lorentz invariant while all on-shell tree amplitudes are automatically twofold gauge invariant.   
We extend our results to curved spacetime and calculate the analogue of the Einstein equations.  While these twofold invariances are hidden in the canonical approach of graviton perturbation theory, they are naturally expected given the double copy relations for scattering amplitudes in gauge theory and gravity.
\end{quote}

\setcounter{footnote}{0}

\newpage
\tableofcontents

\newpage

\section{Introduction}\label{sec:introduction}

The scattering amplitudes program has revealed extraordinary structures underlying long-studied quantum field theories.  One such class of miracles reformulates gravity as the ``square'' of gauge theory.  Discovered by KLT \cite{KLT} and generalized by BCJ \cite{BCJ}, this relationship is encoded in concrete formulae expressing the scattering amplitudes of pure gravity as sums over products of the scattering amplitudes of Yang-Mills theory,
\be
A_{\rm GR} \sim \sum A_{\rm YM} \bar A_{\rm YM},
\label{eq:square}
\ee
where the barred and unbarred factors need not be the same amplitude.
This duality appears in various guises in a variety of contexts, both in field theory and string theory (see \Ref{Carrasco:2015iwa} and refs. therein). Remarkably, the double copy structure also persists in classical field theory, where certain gauge theory backgrounds map directly to solutions of general relativity \cite{Monteiro:2014cda,Ridgway:2015fdl,Luna:2015paa,Luna:2016due,Chu:2016ngc,Goldberger:2016iau}.

Pragmatically, these squaring relations simplify certain gravity calculations by connecting them directly  to known computations in gauge theory \cite{CHY}.  From a top-down perspective, however, this correspondence suggests a very surprising fact about the underlying symmetries of gravity.  In particular, since the right-hand side of \Eq{eq:square} is a sum over products of Lorentz scalars, it is separately invariant under Lorentz transformations acting individually on each Yang-Mills factor.  To see this explicitly, consider graviton polarizations expressed as a bivector,
\be
\epsilon^{\vphantom{}}_{a \bar b} = \epsilon^{\vphantom{}}_a \bar\epsilon_{\bar b},
\ee
where $A_{\rm YM}$ and $\bar A_{\rm YM}$ depend only on $\epsilon$ and $\bar \epsilon$, respectively.  Denoting the momenta contracted with unbarred and barred indices by $k$ and $\bar k$, respectively, it follows that the right-hand side of \Eq{eq:square} exhibits a formal twofold invariance under a pair of Lorentz transformations,
\be 
k^{\vphantom{}}_a \rightarrow \Lambda_a^{\;\;b} k^{\vphantom{}}_b \qquad \textrm{and} \qquad \bar k^{\vphantom{}}_{\bar a} \rightarrow \bar \Lambda_{\bar a}^{\;\;\bar b} \bar k_{\bar b},
\ee
together with a pair of Ward identity transformations,
\be 
\epsilon_a \rightarrow \epsilon_a + k_a \qquad \textrm{and} \qquad \bar \epsilon_{\bar a} \rightarrow \bar \epsilon_{\bar a} + \bar k_{\bar a}. \label{eq:doublegauge}
\ee
The fact that the physical scattering amplitudes of pure gravity can be expressed as products of Yang-Mills amplitudes hints at an underlying ``twofold Lorentz symmetry'' of pure gravity,
\be 
SO(D-1,1) \times \overline {SO}(D-1,1).
\label{eq:SOD2}
\ee
It should be possible to manifest such a property at the level of the action.  Such a formulation would manifest ``index factorization'', i.e., where all interactions of the graviton field $h_{a \bar b}$ involve indices contracted with
$\eta_{ab}$ and 
$\eta_{\bar a \bar b}$, thus forbidding contractions between barred and unbarred indices. 
This condition places stringent restrictions on the allowed interaction terms.  For example, something as innocuous as the trace of the graviton, $h_a^{\;\;a} = h_{a \bar b} \eta^{\bar b a}$,   is not twofold Lorentz invariant since $\epsilon^{\vphantom{}}_a \bar \epsilon^a = \epsilon^{\vphantom{}}_a \bar \epsilon_{\bar b} \eta^{\bar b a}$ contracts barred and unbarred indices.

The canonical procedure for graviton perturbation theory grossly violates index factorization and, in turn, twofold Lorentz symmetry.  In particular, the Einstein-Hilbert (EH) action in $D$ spacetime dimensions is\footnote{We work in mostly-plus signature and use the conventions $R_{ab} = R^c_{\;\;acb}$ and $R^a_{\;\;bcd} = \partial_c \Gamma^a_{bd} -\partial_d \Gamma^a_{bc}+ \Gamma^a_{ce}\Gamma^e_{bd} - \Gamma^a_{de}\Gamma^e_{bc}$. We denote the determinant of a metric as the metric's label with no indices, e.g., $g = \det g_{ab}$, etc. For notational reasons, we will adopt Latin indices throughout.}
\be 
S = \int  \mathrm{d}^D x\, \sqrt{-g} \, \left( \frac{R}{16\pi G}  + \LGF \right)\label{eq:preaction},
\ee
where $\LGF$ denotes the Faddeev-Popov gauge-fixing term.  To compute graviton scattering amplitudes in perturbation theory, we typically define
\be 
g_{ab} =\eta_{ab} + h_{ab}\label{eq:linearized}
\ee
and expand the action in powers of the graviton perturbation $h_{ab}$.  Terms involving the trace of the graviton, together with other nonfactorized structures, induce propagators and interaction vertices that explicitly violate the twofold Lorentz symmetry.  

Nevertheless, in the seminal work of \Ref{Bern:1999ji}, Bern and Grant showed how the KLT relations can be reverse-engineered to perturbatively construct an action for pure gravity compatible with manifest index factorization.  They achieved this feat up to fifth order in graviton perturbations, leaving open the question of an all-orders generalization.  Furthermore, to derive this action from the original EH action required the introduction of a dilaton, which when integrated out induced nonlocal graviton interactions.

In this paper, we recast the EH action for pure gravity into a form that is local and manifestly twofold Lorentz invariant at all orders in graviton perturbations.  To do so, we exploit the fact that the usual EH action of conventional graviton perturbation theory is not particularly meaningful: the freedom of nonlinear field redefinitions and gauge fixing permits one to rewrite the action in an infinite number of different ways, all describing equivalent physics.   By exploring this full freedom, we derive a local representation of the EH action that is compatible with index factorization at all nonlinear orders and requires no additional dynamical or auxiliary fields beyond the graviton.  The off-shell Feynman propagators and vertices  are trivially twofold Lorentz invariant and the resulting tree-level on-shell scattering amplitudes are twofold gauge invariant.  The resulting action is derived most naturally in an ``exponential basis'' for the graviton, reminiscent of the common parameterization of Nambu-Goldstone bosons in the nonlinear sigma model.

By recasting this action in terms of fields on a doubled spacetime of dimension $2D$, we automate the bookkeeping of the barred and unbarred indices at the expense of introducing a two-form field, which decouples from all tree-level graviton scattering amplitudes.  We comment on the link between these representations and those that arise from double field theory \cite{Hull:2009mi,Hohm:2010jy,Hohm:2010pp,Hohm:2011dz,Siegel:1993xq,Siegel:1993th,Siegel:1993bj}, where Einstein gravity coupled to a dilaton and two-form arises as the low-energy effective field theory of string theory at leading order in the derivative expansion. 

An obvious corollary is that our action also generates, via further field redefinitions, an infinite class of equivalent twofold Lorentz invariant actions.   Again utilizing this freedom of graviton field basis, we study alternative versions of this action, going from the exponential basis to the analogue of the ``Cayley basis'' \cite{Kampf:2013vha} for the nonlinear sigma model.  Here, we find that graviton perturbation theory simplifies substantially and manifests some unexpected additional symmetries. 

The remainder of this paper is organized as follows.
In \Sec{sec:FD}, we discuss a systematic procedure parameterizing the space of local field redefinitions and gauge-fixing conditions in pure gravity.  Afterwards, we show in \Sec{sec:factor} how this exercise yields a simple action that exhibits index factorization and thus twofold Lorentz invariance.  This form is naturally written in terms of a spacetime of doubled dimension.  We then discuss the graviton propagator, as well as a more general class of twofold Lorentz invariant theories related by field redefinitions.    Next, we generalize this formalism to curved spacetime in \Sec{sec:curved}, establishing index factorization for any Ricci-flat spacetime and deriving the corresponding Einstein equations.  We conclude and discuss future directions in \Sec{sec:conclusions}.

\section{Building the Action}\label{sec:FD}

In this section, we define the space of local actions equivalent to the EH action modulo field redefinitions and gauge fixing.   For a particular choice, the EH action can be recast into a form that manifests index factorization and is thus compatible with twofold Lorentz invariance.  Here, we will study graviton perturbation theory as an expansion about flat spacetime in Cartesian coordinates,
\be 
\eta_{ab} = \mathrm{diag}(-1,1,\ldots,1).
\ee
In \Sec{sec:curved}, we will generalize our results to arbitrary backgrounds and curvilinear coordinate systems.

\subsection{Index Factorization}

To begin, we identify which terms are compatible and incompatible with index factorization.  
For later convenience, we define powers of the graviton tensor by
\be 
\begin{aligned}
h^n_{ab} &= h_{a  b_1} \eta^{b_1 a_1}h_{a_1 b_2}  \eta^{b_2 a_2}  \cdots  h_{a_{n-2} b_{n-1}}  \eta^{b_{n-1} a_{n-1}}h_{a_{n-1} b}  \\
&= h_a^{\;\; a_1} h_{a_1}^{\;\;a_2}  \cdots   h_{a_{n-2}}^{\;\; a_{n-1}}h_{a_{n-1} b}, \label{eq:hn}
\end{aligned}
\ee
together with a shorthand for the trace,
\be 
[h^n] = h^n_{ab} \eta^{ba}.
\ee
We can now determine when these products of the graviton tensor are compatible with index factorization.  Many operators are comprised of gravitons built from objects of the form
\be 
[h^{2n}] = \textrm{even cycle}
 \qquad \textrm{or} \qquad
[h^{2n+1}] = \textrm{odd cycle}, \label{eq:cycles}
\ee
where we have suppressed all derivatives and their contractions.
The odd cycles necessarily violate index factorization.  This is obvious because an odd number of graviton tensors appear with an odd number of barred indices and an odd number of unbarred indices.  Thus, contracting all the indices will necessarily involve the contraction of at least one barred and one unbarred index.    In contrast, the even cycles are compatible with index factorization, since there exists a consistent assignment of barred and unbarred indices.

As noted before, however, odd cycles appear ubiquitously in the conventional approach to graviton perturbation theory, which is derived by expanding the EH action in the field basis in \Eq{eq:linearized}.  For example, the volume element is given by
\be
\sqrt{-g} =  \exp\left(\frac{1}{2}\sum_{n=1}^\infty \frac{(-1)^{n-1}}{n}[h^n]\right),\label{eq:detg}
\ee
which has an infinite number of odd cycles that are incompatible with index factorization.  Hence, to construct a representation with manifest index factorization it is necessary to go beyond the standard prescription.  To do so, we rewrite the EH action in an arbitrary local graviton field basis and gauge-fixing, which we now discuss.

\subsection{Field Basis and Gauge Fixing}

To construct an arbitrary field basis, we consider all possible local field redefinitions of the graviton defined in \Eq{eq:linearized}.  
Due to a theorem of Haag \cite{Haag:1958vt} (see also \Ref{Donoghue:1992dd} and refs. therein), field redefinitions leave all scattering amplitudes invariant, provided the asymptotic states remain unaltered.   For example, the local field redefinition of a scalar,
\be 
\phi \rightarrow \alpha_1 \phi + \alpha_2 \phi^2 + \alpha_3 \phi^3 +\cdots,
\ee
leaves scattering amplitudes unchanged provided $\alpha_1 =1$ so that the linearized field is the same.   For the graviton, the analogous field redefinition is
\be 
\begin{aligned}
h_{ab}  \;\;  \rightarrow \quad & \alpha_1 h_{ab} + \alpha_2 \eta_{ab}[h]    \\  
+&  \alpha_3 h^2_{ab} + \alpha_4 h_{ab} [h]+ \alpha_5 \eta_{ab}[h^2]+ \alpha_6 \eta_{ab}[h]^2   \\
+&  \alpha_7 h^3_{ab} + \alpha_8 h^2_{ab} [h]+ \alpha_9h_{ab}[h^2]+ \alpha_{10} h_{ab}[h]^2 + \alpha_{11}  \eta_{ab}[h^3]+ \alpha_{12}  \eta_{ab}[h^2][h] + \alpha_{13} \eta_{ab}[h]^3 \\
+&  \cdots,  \label{eq:FD}
\end{aligned}
\ee 
where $\alpha_1=1$.  Here we will restrict to field redefinitions without any derivatives in order to  maintain the familiar two-derivative form of the graviton interactions.
In general, it is straightforward but tedious to enumerate the various tensor structures at higher orders in the graviton. At $\mathcal{O}(h^n)$, there are $\sum_{j=0}^n p(j)$ possible terms in the nonlinear field redefinition, where $p(j)$ is the number of partitions of $j$.

Next, we consider gauge fixing, which also comes with an immense freedom.  Using the Faddeev-Popov gauge-fixing procedure, we define
\be 
\LGF = -\eta^{ab} F_a F_b,
\ee
for a local but otherwise arbitrary gauge-fixing vector,
\be 
\begin{aligned}
F_a = \partial^b h^{cd} (&
 \beta_1 \; \eta_{ab} \eta_{cd} + \beta_2 \; \eta_{ac} \eta_{bd} + {}  \\
  &   \beta_3 \; h_{ab} \eta_{cd} + \beta_4 \; h_{ac} \eta_{bd}  + \beta_5 \; \eta_{ab} h_{cd} + \beta_6 \; \eta_{ac} h_{bd} +\beta_7 \; \eta_{ab} \eta_{cd}[h] + \beta_8 \; \eta_{ac} \eta_{bd}[h] + \cdots ),\label{eq:GF} 
\end{aligned}
\ee
which can be thought of as a highly nonlinear generalization of harmonic gauge. At $\mathcal{O}(h^n)$ in the nonlinear gauge-fixing vector, there are $2\sum_{j=1}^n j\;p(n-j)$ possible terms.

As noted earlier, the $\alpha$ and $\beta$ parameters that appear in the field basis and gauge-fixing have absolutely no effect on physical scattering amplitudes.  However, as a check of our calculation, we have also explicitly computed the three-particle and four-particle scattering amplitudes and verified that they are indeed independent of $\alpha$ and $\beta$.

\section{Factorizing the Action}\label{sec:factor}

The $\alpha$ and $\beta$ parameters of the field basis and gauge-fixing alter the action but have no effect on physical observables.  Next, we can examine the action at each order in graviton perturbations, fixing the $\alpha$ and $\beta$ parameters so as to precisely eliminate all appearances of odd cycles, as defined in \Eq{eq:cycles}.  This is a necessary condition for manifest index factorization.  By explicit computation, we have verified that this criterion can be satisfied at least up to fifth order in the graviton.  Perhaps surprisingly, we have also found that a special choice of the $\alpha$ and $\beta$ parameters follows a simple pattern that straightforwardly generalizes to all orders in perturbation theory, taking a simple analytic form. 
One can then prove that this choice of nonlinear field redefinition and gauge-fixing allows for index factorization of the action at {\it all orders} in the graviton. It is to this special class of field redefinitions and gauge-fixing that we now turn.

\subsection{Definition of the Action}

We focus on a special field basis for the graviton defined by 
\be 
g_{ab} = \eta_{ab} + \pi_{ab} + \frac{1}{2!} \pi^2_{a b}+ \frac{1}{3!} \pi^3_{ab}+\cdots, \qquad \textrm{where} \qquad
\pi_{ab} = h_{ab} - \frac{1}{D-2} \eta_{ab} [h].\label{eq:pi} 
\ee
It will often be convenient to invoke the shorthand notation
\be 
g_{ab} = (e^\pi)_{ab}  \qquad \textrm{and} \qquad g^{ab} = (e^{-\pi})^{ab},\label{eq:FDexp}
\ee
where by construction $g_{ab} g^{bc} = \delta_a^c$.  We emphasize here that $g_{ab}$ and $g^{ab}$ are matrix inverses, not related by raising and lowering with respect to $\eta_{ab}$. The utility of an exponential basis for gravity, in that it treats the metric and its inverse symmetrically in the perturbation expansion, was understood previously in \Ref{ElvangNotes}.
Our Faddeev-Popov gauge-fixing term is
\be 
\LGF = -\frac{1}{64\pi G(D-2)} e^{[h]/(D-2)}(e^{-h})^{ab} \partial_a [h] \partial_b [h] .
\ee
Using \Eqs{eq:pi}{eq:FDexp}, we see that we can write the gauge-fixing term in the compact form
\be 
\LGF = - \frac{D-2}{64\pi G}  g^{ab} \omega_a \omega_b , \label{eq:LGF}
\ee
where we have defined the vector
\be 
\omega_a = \partial_a \log \sqrt{-g} = -\frac{1}{D-2} \partial_a [h].
\ee
We will postpone further discussion of the physical meaning of this gauge condition to \Sec{sec:curved}.  For now, let us simply view $\LGF$ in \Eq{eq:LGF} as a particular choice of the coefficients in the general gauge-fixing vector in \Eq{eq:GF}.  However, note that the above gauge-fixing term does not eliminate the full gauge freedom: the propagator is not yet invertible, although we will see in \Sec{sec:propagators} how this is remedied by an additional gauge fixing.

Putting everything together, we find that EH action in \Eq{eq:preaction} is drastically simplified, in part because derivatives act nicely on the exponential form of \Eq{eq:pi}.
The resulting action is independent of the spacetime dimension $D$ and can be written compactly as
\be
S = \frac{1}{16\pi G}\int \mathrm{d}^D x \,\partial_a \sigma_{c e} \partial_b \sigma^{de} \left( \frac{1}{4} \sigma^{ab} \delta^c_d- \frac{1}{2} \sigma^{cb}\delta^a_d \right),\label{eq:action}
\ee
expressed in terms of a new exponential field,
\be 
\sigma_{ab} = \eta_{ab} + h_{ab} + \frac{1}{2!} h^2_{a b}+ \frac{1}{3!} h^3_{ab}+\cdots,\label{eq:sigma}
\ee
which we will often express in the shorthand
\be 
\sigma_{ab} = (e^h)_{ab}  \qquad \textrm{and} \qquad \sigma^{ab} = (e^{-h})^{ab},
\ee
where $\sigma_{ab}\sigma^{bc}= \delta_a^c$.  Note that to obtain \Eq{eq:action} we applied the useful identity $\sigma^{ab}\partial_c \sigma_{ab} = \partial_c [h]$, valid in Cartesian coordinates so the metric has unit determinant. 

\Eq{eq:action} is a primary result of this paper, so let us pause to discuss some salient points.  First, since we derived this action directly from the EH action, it is a completely equivalent description of pure gravity expanded around flat spacetime.  Consequently, the scattering amplitudes computed with this action are exactly equal to those obtained in conventional graviton perturbation theory.  

Second, \Eq{eq:action} is constructed so that every interaction is compatible with index factorization.  Consequently, it is always possible to assign distinct sets of barred and unbarred indices that are separately contracted.  For example, our field basis is chosen to precisely eliminate the $\sqrt{-g}=e^{-[h]/(D-2)}$ factor, which was a persistent source of odd cycles in the action.  This factor is precisely canceled by the factors of $[h]$ in the definition of $\pi_{ab}$ in \Eq{eq:pi}.   Formally, two sets of independently contracted indices exhibit an enhanced twofold Lorentz symmetry.  However, these are not, at least in this particular form, symmetries in the literal sense because they act as rigid transformations on the barred and unbarred indices, as for, e.g., an internal symmetry.    In terms of the scattering amplitudes relations in \Eq{eq:square}, this restriction of the enhanced symmetry comes from the fact that the two Yang-Mills amplitudes are separately Lorentz invariant, but crucially must have the same external momenta.  As we will soon see, by introducing auxiliary extra dimensions one can promote this property of index factorization into a bona fide symmetry of the action.

Third, it is remarkable how the exponential field defined in \Eq{eq:sigma} arises naturally from our prescription for eliminating odd cycles.  This object is curiously reminiscent of the exponential parameterization of the nonlinear sigma model.  It is tempting to imagine that this form of the EH action implies some form of underlying spontaneous symmetry breaking within gravity.   However, as we will see later, there are many alternative field bases that are not exponential.

Fourth, \Eq{eq:action} is extremely simple compared to the standard action for graviton perturbations, which is derived by inserting the field basis of \Eq{eq:linearized} into \Eq{eq:preaction}.  Expanding \Eq{eq:action} in perturbations, we find that
\be
S = \frac{1}{16\pi G}\int\mathrm{d}^D x\, \sum_n {\cal O}_n,\label{eq:actionexpand}
\ee
where the first few orders of the operators ${\cal O}_n$ are
\begin{align}
{\cal O}_2 =&   + \frac{1}{2}\partial_c h_{ab} \partial^b h^{ac} - \frac{1}{4} \partial_c h_{ab} \partial^c h^{ab} \nonumber \\ 
{\cal O}_3 =&  +\frac{1}{4} h^{ab} \partial_a h_{cd} \partial_b h^{cd} -  \frac{1}{2} h^{ab} \partial_c h_{ad} \partial_b h^{cd} \nonumber \\  
{\cal O}_4 =& +\frac{1}{8} h_{ab} h^{cd} \partial^b h_{ce} \partial_d h^{ae} - \frac{1}{8} h^{ab}h_{ac}\partial_b h_{de} \partial^c h^{de} - \frac{1}{12}h^{ab} h^{cd} \partial_c h_{be} \partial^e h_{ad} +\frac{1}{24}h^{ab}h^{cd}\partial^e h_{cb} \partial_e h_{ad} \qquad \nonumber \\
 &+\frac{1}{6}h^{ab} h_{ac} \partial^c h^{de} \partial_e h_{db}+\frac{1}{24}h^{ab}h_{ac}\partial^d h^{ec} \partial_e h_{db}-\frac{1}{24}h^{ab}h_{ac}\partial^e h^{dc} \partial_e h_{db} \label{eq:EHpert} \\ 
{\cal O}_5 =& -\frac{1}{12} h^{ab}h_{ac}h_{de}\partial^c h^{fe}\partial^d h_{fb} + \frac{1}{24}h^{ab}h_{ac}h_{db}\partial^c h^{ef} \partial^d h_{ef} + \frac{1}{24}h^{ab} h^{cd} h^{ef} \partial_c h_{eb} \partial_f h_{ad} \nonumber \\
&+\frac{1}{24}h^{ab}h_{ac}h^{de}\partial_d h_{fb} \partial_e h^{fc}-\frac{1}{24}h^{ab}h^{cd}h^{ef} \partial_e h_{ad} \partial_f h_{cb} + \frac{1}{24} h^{ab} h_{ac} h^{de} \partial^c h_{fe} \partial^f h_{db} \nonumber 
\\&
-\frac{1}{24} h^{ab} h_{ac} h^{de} \partial_e h^{fc} \partial_f h_{db} - \frac{1}{24} h^{ab}h_{ac}h_{db}\partial^d h_{ef} \partial^f h^{ec}. \nonumber
\end{align}
It is straightforward to check that in all of these interactions it is always possible to assign independent sets of barred and unbarred indices that never contract with one another.  

While \Eq{eq:action} is compatible with index factorization, it is certainly not ideal that checking this requires running through each interaction term one at a time and intelligently assigning barred and unbarred indices.  Indeed, the situation would be substantively improved with a formalism that does not require a case-by-case analysis of each term, instead treating indices as barred and unbarred from the very beginning.  We construct just such a representation in the next subsection. 

\subsection{Adding Auxiliary Dimensions}

To automate the proper contraction of barred and unbarred indices, we introduce an additional set of auxiliary dimensions.  In particular, let us extend the $D$ dimensions of spacetime into $2D$ dimensions, where
\be 
x^A = (x^a,  \bar{x}^{\bar a}) \qquad \textrm{and} \qquad
\partial_A = (\partial_a, \partial_{\bar a})
\ee
and the original $D$-dimensional spacetime corresponds to the restriction to the ``diagonal'' spacetime
\be 
x^a = \bar{x}^{\bar a}.
\ee
Here, indices in $2D$-dimensional spacetime are contracted with the metric tensors
\be 
\eta_{AB} =  \left[
\begin{array}{cc}
\eta_{a b} &0 \\
0 & \eta_{\bar a \bar b}
\end{array}
\right] \quad \textrm{and} \quad 
\eta^{AB} =  \left[
\begin{array}{cc}
\eta^{a b} &0 \\
0 & \eta^{\bar a \bar b}
\end{array}
\right],
\ee
so all terms are automatically twofold Lorentz invariant with respect to barred and unbarred indices.

Next, we repackage the graviton into a tensor in $2D$-dimensional spacetime,
\be 
H_{AB} =  \left[ 
\begin{array}{cc}
0 & h_{a \bar b} \\
h_{\bar a  b} & 0 
\end{array}
\right],
\ee
where the two off-diagonal blocks are transposes of each other.  The structure of this representation explicitly breaks the underlying $SO(2D-2,2)$ symmetry of the doubled $2D$-dimensional spacetime down to the symmetry in \Eq{eq:SOD2}.  Since barred and unbarred indices are distinct, $h_{a \bar b}$ is automatically lifted to a general $D$-dimensional matrix.  The usual physical graviton modes correspond to the symmetric components of this tensor.  As we will see shortly, the antisymmetric component can be neglected at tree level for graviton scattering amplitudes.
In terms of this new field, we define the exponential field
\be 
\Sigma_{AB} = (e^H)_{AB}  \quad \textrm{and} \quad \Sigma^{AB} = (e^{-H})^{AB}.
\ee
A simple computation shows that
\be 
\Sigma_{AB} =\left[
\begin{array}{cc}
(\cosh h)_{ab} & (\sinh h)_{a \bar b}  \\
(\sinh h)_{\bar ab}  & (\cosh h)_{\bar a \bar b}  
\end{array}
\right] ,
\label{eq:coshsinh}
\ee
where, in our shorthand,
\be 
\begin{aligned}
(\cosh h)_{ab} = \eta_{ab} + \frac{1}{2!} h^2_{ab } +  \frac{1}{4!} h^4_{ab} + \cdots \qquad \textrm{and}\qquad
(\sinh h)_{a\bar b} = h_{a \bar b} + \frac{1}{3!} h^3_{a\bar b} + \frac{1}{5!} h^5_{a\bar b}+\cdots
\end{aligned}
\ee
are even and odd functions in the graviton, respectively.  Because these terms have distinct parity, we can, in analogy with \Eq{eq:hn}, define
\be 
\begin{aligned}
h^{2n}_{ab} &= h_{a  \bar b_1} \eta^{ \bar b_1 \bar  a_1}h_{\bar  a_1 b_2}  \eta^{b_2 a_2}  \cdots  h_{a_{2n-2} \bar  b_{2n-1}}  \eta^{\bar  b_{2n-1} \bar  a_{2n-1}}h_{\bar a_{2n-1} b} \\
&= h_a^{\;\; \bar a_1} h_{\bar a_1}^{\;\;a_2}  \cdots   h_{a_{2n-2}}^{\;\; \bar a_{2n-1}}h_{\bar a_{2n-1} b} 
\end{aligned}
\ee
for even powers of the graviton, while for odd powers of the graviton,
\be 
\begin{aligned}
h^{2n+1}_{a\bar b} &= h_{a  \bar b_1} \eta^{ \bar b_1 \bar  a_1}h_{\bar  a_1 b_2}  \eta^{b_2 a_2}  \cdots  h_{\bar a_{2n-1}  b_{2n}}  \eta^{  b_{2n} a_{2n}}h_{ a_{2n} \bar b}  \\
&= h_a^{\;\; \bar a_1} h_{\bar a_1}^{\;\;a_2}  \cdots   h_{\bar a_{2n-1}}^{\;\; a_{2n}}h_{a_{2n} \bar b}
\label{eq:oddgravpower}
\end{aligned}
\ee
and similarly for the other tensors.  By construction, we see that the barred and unbarred indices are never contracted with each other.

In terms of these new variables, the action takes the form
\be 
S = \frac{1}{16\pi G}\int \mathrm{d}^D x \, \mathrm{d}^D \bar{x} \, \delta^D(x-\bar{x}) \, \partial_A \Sigma_{C E} \partial_B \Sigma^{DE} \left( \frac{1}{16} \Sigma^{AB} \delta^C_D- \frac{1}{4} \Sigma^{CB}\delta^A_D \right),\label{eq:actionfactor}
\ee
where the numerical factors are slightly different from those in \Eq{eq:action} due to additional factors of two coming from the trace over the $2D$-dimensional spacetime. Notably, \Eq{eq:actionfactor} has several  properties not manifest in the usual representation of the EH action, which we now discuss.  

First and foremost, the action is manifestly invariant under a twofold Lorentz symmetry that acts separately on $x$ and $\bar x$. Due to the $\delta$ function in \Eq{eq:actionfactor}, i.e., the fact that the action is only integrated over the diagonal combination $x = \bar x$, the two corresponding conserved currents are one and the same.  In particular, they produce the usual single conservation of energy, momentum, and angular momentum in $D$-dimensional spacetime. 
To see the index factorization explicitly, we can again expand the action in perturbations to obtain
\begin{align}
{\cal O}_2 =&  + \frac{1}{4} \partial_c h_{a \bar b} \partial^a h^{c \bar b} +\frac{1}{4} \partial_{\bar c} h_{a \bar b} \partial^{\bar b} h^{a \bar c} -\frac{1}{8} \partial_c h_{a \bar b} \partial^c h^{a\bar b} -\frac{1}{8} \partial_{\bar c} h_{a \bar b} \partial^{\bar c}h^{a\bar b} \nonumber \\[0.5ex]
{\cal O}_3 =& + \frac{1}{4} h^{a \bar b} \partial_a h_{c \bar d} \partial_{\bar b} h^{c \bar d}  - \frac{1}{4} h^{a \bar b} \partial_d h_{a \bar c} \partial_{\bar b} h^{d \bar c} - \frac{1}{4} h^{a\bar b}\partial_{\bar d} h_{c\bar b} \partial_a h^{c\bar d} \nonumber \\[0.5ex]
{\cal O}_4 =& -\frac{1}{16}h^{a \bar b} h_{c \bar b} \partial_a h^{d \bar e} \partial^c h_{d \bar e} - \frac{1}{16} h^{a \bar b} h_{a \bar c} \partial_{\bar b} h^{d \bar e} \partial^{\bar c} h_{d \bar e} + \frac{1}{16} h_{a \bar b} h^{c \bar d} \partial^a h_{e \bar d} \partial_c h^{e \bar b} \nonumber \\&+ \frac{1}{16} h_{a \bar b} h^{c \bar d} \partial^{\bar b} h_{c \bar e} \partial_{\bar d} h^{a \bar e}-\frac{1}{24} h_{a \bar b} h_{c \bar d} \partial^c h^{e \bar b} \partial_e h^{a \bar d} - \frac{1}{24} h_{a \bar b} h_{c \bar d} \partial^{\bar b} h^{c \bar e} \partial_{\bar e} h^{a \bar d} \nonumber \\&+ \frac{1}{48} h^{a \bar b} h^{c \bar d} \partial_e h_{c \bar b} \partial^e h_{a \bar d} + \frac{1}{48} h^{a \bar b} h^{c \bar d} \partial_{\bar e} h_{c \bar b} \partial^{\bar e} h_{a \bar d}+\frac{1}{12} h^{a \bar b} h_{c \bar b} \partial^c h^{e \bar d} \partial_e h_{a \bar d}  \\&+\frac{1}{12} h^{a \bar b} h_{a \bar c} \partial^{\bar c} h^{d \bar e} \partial_{\bar e} h_{d \bar b}-\frac{1}{96}h^{a \bar b} h_{c \bar b} \partial^e h^{c \bar d} \partial_e h_{a \bar d} - \frac{1}{96} h^{a \bar b} h_{c \bar b} \partial^{\bar e} h^{c \bar d} \partial_{\bar e} h_{a \bar d} \nonumber \\
&+\frac{1}{48} h^{a \bar b} h_{a \bar c} \partial^d h^{e \bar c} \partial_e h_{d \bar b} + \frac{1}{48} h^{a \bar b} h_{c \bar b} \partial^{\bar d} h^{c \bar e} \partial_{\bar e} h_{a \bar d} - \frac{1}{96} h^{a \bar b} h_{a \bar c} \partial^e h^{d \bar c} \partial_e h_{d \bar b} \nonumber \\&- \frac{1}{96} h^{a \bar b} h_{a \bar c} \partial^{\bar e} h^{d \bar c} \partial_{\bar e} h_{d \bar b}\nonumber 
\\
{\cal O}_5 =& +\frac{1}{24} h^{a \bar b} h_{a \bar c} h_{d \bar b} \partial^{\bar c} h^{e \bar f} \partial^d h_{e \bar f} - \frac{1}{24} h^{a \bar b} h_{a \bar c} h_{d \bar e} \partial^{\bar c} h^{f \bar e} \partial^d h_{f \bar b} + \frac{1}{24} h^{a \bar b} h^{c \bar d} h^{e \bar f} \partial_{\bar b} h_{c \bar f} \partial_e h_{a \bar d} \nonumber \\
&- \frac{1}{24} h^{a \bar b} h_{c \bar b} h_{d \bar e} \partial^c h^{d \bar f} \partial^{\bar e} h_{a \bar f} + \frac{1}{48} h^{a \bar b} h_{c \bar b} h^{d \bar e} \partial_d h_{a \bar f} \partial_{\bar e} h^{c \bar f} + \frac{1}{48} h^{a \bar b} h_{a \bar c} h^{d \bar e} \partial_d h_{f \bar b} \partial_{\bar e} h^{f \bar c} \nonumber \\
&-\frac{1}{24}h^{a \bar b} h^{c \bar d} h^{e \bar f} \partial_e h_{a \bar d} \partial_{\bar f} h_{c \bar b} + \frac{1}{48} h^{a \bar b} h_{c \bar b} h^{d \bar e} \partial^c h_{d \bar f} \partial^{\bar f} h_{a \bar e} -\frac{1}{48} h^{a \bar b} h_{c \bar b} h^{d \bar e} \partial_d h^{c \bar f} \partial_{\bar f} h_{a \bar e} \nonumber \\
&+\frac{1}{48}h^{a \bar b} h_{a \bar c} h^{d \bar e} \partial^{\bar c} h_{f \bar e} \partial^f h_{d \bar b}-\frac{1}{48} h^{a \bar b} h_{a \bar c} h^{d \bar e} \partial_{\bar e} h^{f \bar c} \partial_f h_{d \bar b} -\frac{1}{48} h^{a \bar b} h_{a \bar c} h_{d \bar b} \partial^{\bar c} h_{f \bar e} \partial^f h^{d \bar e} \nonumber \\&-\frac{1}{48} h^{a \bar b} h_{a \bar c} h_{d \bar b} \partial^d h_{e \bar f} \partial^{\bar f} h^{e \bar c}. \nonumber
\end{align}
As expected, the barred and unbarred indices are all contracted consistently.  

Second, the action \eqref{eq:actionfactor} is manifestly invariant under a $\mathbb{Z}_2$ parity that swaps the two $D$-dimensional spacetimes, 
\be 
\begin{aligned}
x^a  &\leftrightarrow  \bar{x}^{\bar a} \\
h_{a\bar b}  &\leftrightarrow h_{\bar a b}.
\end{aligned}
\ee
  In terms of the full $2D$-dimensional objects, this $\mathbb{Z}_2$ parity acts as
\be 
H  \; \leftrightarrow \;  \tau H \tau \qquad \textrm{and} \qquad
\eta \;  \leftrightarrow \; \tau \eta\tau = \eta,
\ee
where we have defined the swap operator
\be 
\tau_{AB} =  \left[ 
\begin{array}{cc}
0 &\mathbbm{1} \\
\mathbbm{1} & 0 
\end{array}
\right].
\ee
The symmetric and antisymmetric components of $h_{a \bar b}$ are manifestly even and odd under this parity, respectively.  The former corresponds to the usual physical graviton modes, while the latter is an additional two-form field.  However, since the antisymmetric component is odd under the $\mathbb{Z}_2$ parity, it enters the action in pairs and thus does not contribute to tree-level graviton scattering amplitudes.  Thus, since \Eq{eq:actionfactor} is expressed in terms of a general graviton tensor $h_{a \bar b}$, it is, strictly speaking, only equivalent to pure gravity at tree level.   

The above construction is very much reminiscent of one discovered previously in the context of double field theory and there is a close link between our approaches.  Double field theory \cite{Hull:2009mi,Hohm:2010jy,Hohm:2010pp,Hohm:2011dz,Siegel:1993xq,Siegel:1993th,Siegel:1993bj} is derived from the massless modes of closed string field theory on a doubled torus exhibiting a manifest $O(D,D)$ T-duality group.  The resulting low-energy effective theory is comprised of the graviton plus additional massless degrees of freedom: a dilaton and Kalb-Ramond two-form field necessary to maintain  diffeomorphism invariance of the full space.  Similarly motivated by \Ref{Bern:1999ji}, Hohm \cite{Hohm:2011dz} constructed a form of the double field theory action that maintains index factorization as a low-energy remnant of the underlying T-duality.  The resulting action is quite similar to our \Eq{eq:actionfactor}, except that is has both a massless dilaton and two-form.  In this sense, our result is a derivation of a consistent truncation of this action in which the dilaton is not present.
Conversely, the fact that our results are applicable in standard general relativity, i.e., without a dilaton, mean that they are directly relevant for calculations pertinent to our own universe, e.g., scattering amplitudes in Einstein gravity and gravitational wave computations.

\subsection{Scattering Amplitudes}\label{sec:propagators}
The action in \Eq{eq:actionfactor} is a rewriting of the EH action that manifests index factorization and twofold Lorentz symmetry.  We now study how these properties are encoded in scattering amplitudes.  All interaction vertices will be twofold Lorentz invariant even off-shell.  To determine the symmetries of the propagator, we study the kinetic term in momentum space.  Sending $\partial \rightarrow ip$, we obtain
\be 
{\cal O}_2 = \frac{1}{4} h_{a\bar b} h_{c \bar d}  K^{a \bar b c\bar d}, \qquad \textrm{where} \qquad
K^{a \bar b c \bar d} = -p^2 \eta^{ac} \eta^{\bar b \bar d} + \eta^{ac} p^{\bar b} p^{\bar d} + p^a p^c\eta^{\bar b \bar d} .
\ee
We can systematically determine the zero eigenvectors of the kinetic term by solving
\be 
0 =K^{a \bar b c \bar d} h_{c \bar d} = -p^2 h^{a \bar b} + h^{a\bar d} p_{\bar d} p^{\bar b} +  p^a p_c h^{c\bar b},
\label{eq:eigenmode}
\ee
where indices are raised and lowered with $\eta_{ab}$ and $\eta_{\bar a \bar b}$.
Dotting this equation into $p_a$, we obtain
\be 
0 = p_a h^{a \bar d} p_{\bar d} p^{\bar b},
\ee
which is trivially satisfied for the antisymmetric component of $h_{a\bar b}$.  This equation also vanishes for the symmetric component of $h_{a \bar b}$ when it takes the form of a transverse diffeomorphism, $h_{a b} = \partial_a \xi_b + \partial_b \xi_a$ with
$\partial_b \partial_a \xi^a = 0$.
The existence of zero eigenmodes of the kinetic term implies that $h_{a \bar b}$ does not yet have an invertible kinetic term.

To remedy this, recall that the antisymmetric component of $h_{a \bar b}$ enters the action in pairs on account of the underlying $\mathbb{Z}_2$ parity, so it decouples from tree-level graviton scattering.    We must also, however, modify the gauge-fixing of the symmetric piece in order to produce an invertible kinetic term.  In principle, there are many prescriptions for doing so.  Here we consider a gauge-fixing that is manifestly twofold Lorentz symmetric at the expense of the $\mathbb{Z}_2$ parity, so
\be 
K_{\xi}^{a \bar b c \bar d} = -p^2 \eta^{ac} \eta^{\bar b \bar d} +  \left(1+ \frac{1}{\xi}\right)\eta^{ac} p^{\bar b} p^{\bar d} + \left(1- \frac{1}{\xi}\right) p^a p^c \eta^{\bar b \bar d} ,\label{eq:Kxi}
\ee
where we take $\xi \rightarrow 0$ in the analogue of Landau gauge for gauge theory. 
The corresponding propagator, $\Delta_{a \bar b c \bar d}$, satisfies
\be 
K_{\xi}^{a \bar b c \bar d}  \Delta_{c \bar d e \bar f} = i\delta^a_e \delta^{\bar b}_{\bar f},
\ee
from which we obtain
\be 
 \Delta_{a \bar b c \bar d} = -\frac{i}{p^2} \left( \eta_{ac}\eta_{\bar b \bar d} - (1+\xi) \eta_{ac} \frac{p_{\bar b} p_{\bar d}}{p^2}- (1-\xi)  \frac{ p_{a} p_{c}}{p^2}   \eta_{\bar b \bar d }\right). \label{eq:prop}
\ee
At zeroth order in $\xi$, the $\mathbb{Z}_2$ parity of the propagator is restored, yielding a simple and convenient propagator for explicit computations.  Contributions first order in $\xi$ also vanish because the underlying $\mathbb{Z}_2$ parity of the interactions eliminates all odd powers of $\xi$ dependence from the tree-level graviton scattering amplitude. Note that to obtain consistent answers, it is crucial to use the fully gauge-fixed propagator in \Eq{eq:prop} with the the factorized action in \Eq{eq:actionfactor} and its perturbative expansion.  That is, dropping the delineation between barred and unbarred indices will yield inconsistent results.  In this sense, the two-form is critical for the gauge-fixing introduced in \Eq{eq:Kxi}, even though it does not appear as an external state in graviton scattering amplitudes.  We have checked explicitly that the Feynman diagrams constructed from the propagator in \Eq{eq:prop} and the interaction vertices of \Eq{eq:actionfactor} produce the correct three-, four-, and five-point amplitudes, even for finite $\xi$.

More generally, in this gauge, all off-shell Feynman diagrams are invariant under twofold Lorentz symmetry as well as $\mathbb{Z}_2$ exchange.  Furthermore, the resulting tree-level scattering amplitudes are invariant under the twofold Ward identities defined in \Eq{eq:doublegauge}.  The reason for this is simple: the symmetric combination of gauge transformations is an invariance of graviton scattering, while the antisymmetric combination decouples because this mode only enters in pairs and thus does not contribute to pure graviton scattering at tree level.

Finally, let us emphasize that the action presented here is distinct from the action constructed perturbatively up to fifth order in \Ref{Bern:1999ji}.  This is evident from our propagator, which is different from the propagator assumed in \Ref{Bern:1999ji}.  

\subsection{Alternative Representations}\label{sec:simplicity}

We have presented a simple representation of the EH action that manifests index factorization and in turn twofold Lorentz symmetry.  Now, by again exploiting the freedom to choose a field basis, we can generate an infinite class of physically equivalent actions that manifest the same symmetries.  In particular, we can consider field redefinitions of the form
\be 
h_{a\bar b} \rightarrow \alpha_1 h_{a\bar b} +  \alpha_3 h^{3}_{a\bar b}  + \alpha_5 h^{5}_{a\bar b} +\cdots,\label{eq:FDfurther}
\ee
where again we have $\alpha_1=1$ to maintain the form of the asymptotic states. Here, the field redefinition involves only odd powers of the graviton defined by \Eq{eq:oddgravpower}, so that barred and unbarred indices are properly contracted.  More generally, one can consider an arbitrary sum over $h^n_{a \bar b}$ for odd $n$, with each term multiplied by $[h^m]$ for some even $m$, which preserves the ability to consistently factorize indices. 

In general, this additional set of field redefinitions can further simplify various parts of the action.  For example, to eliminate the appearance of hyperbolic functions in \Eq{eq:coshsinh}, we could send
\be 
h_{a\bar b} \rightarrow (\sinh^{-1} h)_{a \bar b} = h_{a \bar b} - \frac{1}{6} h^{3}_{a \bar b} + \frac{3}{40}h^{5}_{a \bar b} + \cdots,
\ee
so that the EH action is just as in \Eq{eq:actionfactor}, except with a new field defined as
\be 
\Sigma_{AB} \rightarrow \left[
\begin{array}{cc}
(\sqrt{1+ h^2})_{ab} & h_{a \bar b}  \\
h_{\bar ab}  & (\sqrt{1+ h^2})_{\bar a \bar b}  
\end{array}
\right] .
\ee 
In what follows, we discuss an alternative field basis for the action that results in even simpler expressions for graviton perturbation theory.

In particular, inspired by the so-called Cayley basis for the nonlinear sigma model action \cite{Kampf:2013vha}, it is natural to consider the field redefinition 
\be
h_{a\bar b} \rightarrow \log  \left(\frac{1 + \frac{1}{2}h}{1 - \frac{1}{2}h }\right)_{a \bar b}  =  h_{a \bar b} + \frac{1}{12} h^{3}_{a \bar b} + \frac{1}{80} h^{5}_{a \bar b} + \cdots,\label{eq:Cayley}
\ee
for which the field in the doubled spacetime becomes
\be 
\Sigma_{AB} =\left[ 
\begin{array}{cc}
\left(\frac{1+h^2/4}{1-h^2/4}\right)_{ab} & \left(\frac{h}{1-h^2/4}\right)_{a \bar b}  \\
\left(\frac{h}{1-h^2/4}\right)_{\bar ab}  & \left(\frac{1+h^2/4}{1-h^2/4}\right)_{\bar a \bar b}
\end{array}
\right].
\ee
The first few terms in the perturbation expansion are
\be
\begin{aligned}
{\cal O}_2 =& + \frac{1}{2}\partial_c h_{ab} \partial^b h^{ac} - \frac{1}{4}\partial_c h_{ab} \partial^c h^{ab}\\
{\cal O}_3 =& + \frac{1}{4} h^{ab} \partial_a h_{cd} \partial_b h^{cd} - \frac{1}{2} h^{ab} \partial_c h_{ad} \partial_b h^{cd}\\
{\cal O}_4 =& + \frac{1}{8} h_{ab} h^{cd} \partial^b h_{ce} \partial_d h^{ae} -\frac{1}{8}h^{ab}h_{ac} \partial_b h_{de} \partial^c h^{de} + \frac{1}{4} h^{ab} h_{ac} \partial^c h^{de} \partial_e h_{db}\\
& +\frac{1}{8} h^{ab} h_{ac} \partial^d h^{ec} \partial_e h_{db} - \frac{1}{8} h^{ab} h_{ac} \partial^e h^{dc} \partial_e h_{db}\\
{\cal O}_5 =& -\frac{1}{8} h^{ab} h_{ac} h_{de} \partial^c h^{fe} \partial^d h_{fb} + \frac{1}{16} h^{ab} h_{ac} h_{db} \partial^c h^{ef} \partial^d h_{ef} + \frac{1}{8} h^{ab} h_{ac} h^{de} \partial_d h_{fb} \partial_e h^{fc}\\
&-\frac{1}{8}h^{ab}h_{ac}h^{de} \partial_e h^{fc} \partial_f h_{db}-\frac{1}{8}h^{ab}h_{ac}h_{db}\partial^d h_{ef} \partial^f h^{ec},
\end{aligned}
\ee
after dropping the distinction between barred and unbarred indices.  

We immediately note that the Cayley-like basis yields fewer terms than our action in \Eq{eq:action}---for which the first few orders are given in \Eq{eq:EHpert}---and far fewer terms than occur in the canonical graviton perturbation theory of the EH action. In particular, at ${\cal O}(h^n)$ for $n = 2,3,4,5$, the canonical graviton perturbation yields 4, 13, 35, 76 terms in the action, respectively, counted such that no single graviton is acted upon with two derivatives.

A unique aspect of the Cayley-like basis \eqref{eq:Cayley} is that it makes the action invariant up to a sign-flip under the duality of small and large graviton perturbations. Specifically, consider a metric perturbation $h_{ab}$ that has a nonsingular matrix inverse $h^{-1}_{ab}$. Then, in the Cayley-like basis, the transformation 
\be 
\frac{h_{ab}}{2}  \rightarrow \left( \frac{h_{ab}}{2} \right)^{-1}
\ee 
merely induces a sign in the field
\be 
\sigma_{ab} \rightarrow -\sigma_{ab}
\ee
and thus sends the action to minus itself, which simply flips the sign of $\hbar$ and is thus an invariance of the interactions.  This invariance, which is unique to the Cayley-like basis, is reminiscent of T-duality, but is more general in the sense that it applies to arbitrary invertible metric perturbations, while more specific in that it applies to the pure gravity theory considered in this paper. 

\section{Generalizing to Curved Spacetime}\label{sec:curved}

In \Secs{sec:FD}{sec:factor}, we presented a factorized form of the pure gravity action expanded around a flat background. We will now generalize this construction to curved spacetime, first in terms of the full metric and then for perturbations around a nontrivial background.  Afterwards, we derive the corresponding factorized Einstein equations.

\subsection{Lifting to Curved Spacetime} 

Although the action in \Eq{eq:action} was derived by expanding about flat spacetime, it remains valid to all orders in the graviton perturbation.  This implies that this action encodes the physics of large graviton field variations away from flat spacetime, i.e., a curved background.  In particular, by combining \Eq{eq:pi} with \Eq{eq:sigma}, we see that the nonlinear field defined earlier is simply
\be 
\sigma^{ab} = \sqrt{-g} \, g^{ab}. \label{eq:sigma_g}
\ee
Remarkably, this combination of fields arises naturally from the EH action in curved spacetime.  After some rearrangement, one can show that 
\be
\begin{aligned}
\sqrt{-g}\, R &= \sqrt{-g} \left[ \partial_a g_{ce} \partial_b g^{de} \left( \frac{1}{4} g^{ab} \delta^c_d- \frac{1}{2} g^{cb}\delta^a_d \right) - g^{ab} \partial_a \partial_b (\log \sqrt{-g}) \right]+ \text{total derivative}
\\ &= \sqrt{-g} \left[  \partial_a \left(\frac{g_{c e}}{\sqrt{-g}}\right) \partial_b \left(\sqrt{-g} \, g^{de}\right) \left( \frac{1}{4}  g^{ab} \delta^c_d- \frac{1}{2} g^{cb}\delta^a_d \right) \right. \\ &\qquad \qquad  \left. + \frac{D-2}{4} g^{ab} \partial_a (\log \sqrt{-g}) \partial_b (\log \sqrt{-g}) \right]+ \text{total derivative},\label{eq:Ridentity}
\end{aligned}
\ee
which is naturally a function of $\sigma_{ab}$ and $\sigma^{ab}$.  
Something similar arises when we expand in graviton perturbations around a background spacetime $\tilde{g}_{ab}$.  To see this, we lift the nonlinear field into curved spacetime, defining
\be 
\sqrt{-\tilde{g}} \, \sigma^{ab} = \sqrt{-g} \, g^{ab}.\label{eq:sigmaprops}
\ee
Furthermore, we define $\omega_a$ as before and $\tilde{\omega}_a$ analogously,
\be 
 \omega_a = \partial_a \log \sqrt{-g} \qquad \textrm{and} \qquad \tilde{\omega}_a = \partial_a \log \sqrt{-\tilde{g}},
 \ee
as well as their difference,
\be
\Omega_a = \omega_a - \tilde{\omega}_a,
\ee
which enters the curved-background generalization of \Eq{eq:LGF},
\be 
\mathcal{L}_\mathrm{GF} = -\frac{D-2}{64\pi G}g^{ab}\Omega_a \Omega_b.\label{eq:GFcurved} 
\ee
Let us comment on the physical interpretation of this gauge-fixing.
At the level of the gravity action, the gauge condition is a constraint on the full metric $g_{ab}$ or, equivalently, on the metric perturbation $h_{ab}$ in a given field basis. 
The gauge-fixing Lagrangian $\LGF$ itself can be viewed as being added simply to cancel expressions in the non-gauge-fixed equations of motion that vanish when the gauge condition is satisfied. In our case, the gauge condition associated with $\LGF$ is 
\be 
\Omega_a = \partial_a \log\sqrt{\frac{-g}{-\tilde g}} = 0\label{eq:GC},
\ee
which is different from the commonly used harmonic gauge condition, $\partial_b (g^{ab}\sqrt{-g}) = 0$. A gauge condition on the metric can be recast as a condition on the choice of coordinate system $x^a$, regarded as a set of $D$ scalar functions on spacetime. In harmonic gauge, this corresponds to $\nabla_b \nabla^b x^a = 0$. The coordinate condition corresponding to our gauge condition in \Eq{eq:GC}, in terms of the coordinates $x^a$ for the spacetime $g_{ab}$ and $\tilde{x}^a$ for the background spacetime $\tilde{g}_{ab}$, is
\be
\nabla_b \nabla_a x^b = \tilde{\nabla}_b \tilde{\nabla}_a \tilde{x}^b ,\label{eq:GFcoords}
\ee
using that $\nabla_b \nabla_a x^b = - \omega_a$.
Here, $\tilde{\nabla}_a$ is the covariant derivative on the background metric $\tilde{g}_{ab}$ and $\nabla_a$ is the covariant derivative defined with respect to the full perturbed metric $g_{ab}$. 

Armed with the necessary definitions, we are ready to write the gravity action in terms of our field redefinition and gauge-fixing, generalized to an arbitrary background spacetime. First, we note that \Eqs{eq:Ridentity}{eq:GFcurved} imply that \Eq{eq:preaction} is, up to a total derivative,
\be 
S =  \frac{1}{16\pi G} \int \mathrm{d}^D x\,\sqrt{-\tilde{g}}\left[\partial_a \sigma_{ce} \partial_b \sigma^{de} \left(\frac{1}{4}\sigma^{ab}\delta^c_d - \frac{1}{2}\sigma^{cb}\delta^a_d\right) - \sigma^{ab}\partial_a \tilde{\omega}_b\right].
\ee
A useful identity for this simplification is $\sigma^{ab} \partial_c \sigma_{ab} = (2-D)\omega_c + D \tilde{\omega}_c$, which makes use of the fact that $g^{ab}\partial_c g_{ab} = 2\omega_c$. To derive an expression that is manifestly covariant with respect to the background spacetime, we recast partial derivatives in terms of covariant derivatives and Christoffel symbols of the background metric. We then obtain an action that is a nice generalization of \Eq{eq:action} to an arbitrary curved background spacetime,
\be
S = \frac{1}{16\pi G}\int \mathrm{d}^D x \,  \sqrt{-\tilde{g}}\left[\tilde{\nabla}_a \sigma_{ce} \tilde{\nabla}_b \sigma^{de} \left(\frac{1}{4}\sigma^{ab}\delta^c_d - \frac{1}{2}\sigma^{cb}\delta^a_d\right) + \sigma^{ab} \tilde{R}_{ab}\right],\label{eq:actioncurved}
\ee
where $\tilde{R}_{ab}$ is the Ricci tensor of the background spacetime. This action reverts back to \Eq{eq:action} in the flat-spacetime limit.

Note that \Eq{eq:actioncurved} applies independently of the precise field basis for the graviton perturbations, merely requiring the existence of an object $\sigma_{ab}$ consistent with \Eq{eq:sigmaprops}, as well as  the gauge fixing in \Eq{eq:GFcurved}. For concreteness, we now give an explicit field basis for the graviton, for which the required object exists and thus for which \Eq{eq:actioncurved} is the action. Lifting \Eqs{eq:FD}{eq:pi} to curved spacetime, we define the full metric $g_{ab}$ to be
\be 
g_{ab} = \tilde g_{ab} + \pi_{ab} + \frac{1}{2!} \pi^2_{a b}+ \frac{1}{3!} \pi^3_{ab}+\cdots, \qquad \textrm{where} \qquad
\pi_{ab} = h_{ab} - \frac{1}{D-2} \tilde g_{ab} [h]\label{eq:FDcurved} 
\ee
and where we have defined
\be 
\begin{aligned}
h^n_{ab} &= h_{a  b_1} \tilde g^{b_1 a_1}h_{a_1 b_2}  \tilde g^{b_2 a_2}  \cdots  h_{a_{n-2} b_{n-1}}  \tilde g^{b_{n-1} a_{n-1}}h_{a_{n-1} b}  \\
&= h_a^{\;\; a_1} h_{a_1}^{\;\;a_2}  \cdots   h_{a_{n-2}}^{\;\; a_{n-1}}h_{a_{n-1} b},
\end{aligned}
\ee
with traces $[h^n] = h^n_{ab} \tilde g^{ba}$. We now define an exponential field
\be 
\sigma_{ab} = \tilde g_{ab} + h_{ab} + \frac{1}{2!} h^2_{a b}+ \frac{1}{3!} h^3_{ab}+\cdots = (e^h)_{ab} \label{eq:sigmacurved}
\ee
and similarly redefine $\sigma^{ab} = (e^{-h})^{ab}$ with indices in its expansion contracted using $\tilde{g}_{ab}$. With these definitions, along with the useful relation $\sqrt{-g} = \sqrt{-\tilde{g}}\, e^{-[h]/(D-2)}$, the nonlinear field $\sigma_{ab}$ satisfies the property desired in \Eq{eq:sigmaprops}. Hence, the action for the graviton, to all orders in perturbation theory, expanded about an arbitrary background spacetime as in \Eq{eq:FDcurved} and gauge-fixed according to \Eq{eq:GFcurved}, is given in \Eq{eq:actioncurved}. In this field basis, the gauge condition \eqref{eq:GC} becomes $\partial_a [h] = 0$.

Generically, a nonzero value of $\tilde{R}_{ab}$ will violate index factorization since $\sigma^{ab}\tilde{R}_{ab}$ unavoidably contracts left and right indices in all odd powers of $h_{ab}$ in $\sigma^{ab}$.  For example, this will occur in \mbox{(anti-)de Sitter} space, where $\tilde{R}_{ab} \propto \tilde{g}_{ab}$.  However, for a background vacuum solution $\tilde{R}_{ab} = 0$, the action \eqref{eq:actioncurved} factorzies when expressed in terms of $2D$-dimensional spacetime, so
\be 
S = \frac{1}{16\pi G}\int \mathrm{d}^D x \, \mathrm{d}^D \bar{x} \, \delta^D(x-\bar{x}) \, \sqrt{-\tilde{g}} \, \tilde{\nabla}_A \Sigma_{C E} \tilde{\nabla}_B \Sigma^{DE} \left( \frac{1}{16} \Sigma^{AB} \delta^C_D- \frac{1}{4} \Sigma^{CB}\delta^A_D \right),\label{eq:actionfactorcurved}
\ee
where $\tilde{\nabla}_A = (\tilde{\nabla}_a,\tilde{\nabla}_{\bar{a}})$. This action applies for any background vacuum solution to the Einstein equations, including the Schwarzschild and Kerr metrics, Taub-NUT space, gravitational wave backgrounds, etc. In all of these cases, \Eq{eq:actionfactorcurved} provides an all-orders factorized representation of the perturbation theory. As a special case, \Eq{eq:actionfactorcurved} can accommodate any background metric on Minkowski spacetime, e.g., curvilinear coordinates, as opposed to the strict Cartesian coordinate system necessary for the formulation in \Eq{eq:actionfactor}. In addition to the nice factorization properties, the action is very simple in perturbation theory; indeed, the slow scaling of the number of terms discussed in \Sec{sec:simplicity} is equally applicable to \Eq{eq:actionfactorcurved}. Hence, our result may have applicability to the treatment of black hole perturbations, nonlinear gravitational wave effects, etc.

In general, a nontrivial background energy-momentum tensor $\tilde T_{ab}$ will source the Ricci curvature $\tilde{R}_{ab}$, thus violating index factorization.  However, it is simple to see that one particular matter source actually remains compatible with twofold Lorentz symmetry: a massless, minimally-coupled free scalar. For a background source $\tilde{\phi}$, the energy-momentum tensor and Ricci tensor are
\be 
\tilde{T}_{ab} = \partial_a \tilde \phi \partial_b \tilde \phi - \frac{1}{2}\tilde{g}_{ab} \partial_c \tilde \phi \partial^c \tilde \phi \qquad \textrm{and} \qquad 
\tilde{R}_{ab} = 8\pi G\partial_a \tilde \phi \partial_b \tilde \phi.
\ee
 Moreover, the matter action, in the full perturbed spacetime metric with the $\tilde \phi$ background, is 
 \be 
 S_\mathrm{matt} = -\frac{1}{2} \int \mathrm{d}^D x \, \sqrt{-g}\, g^{ab} \partial_a \tilde \phi \partial_b \tilde \phi = -\frac{1}{2} \int \mathrm{d}^D x\,\sqrt{-\tilde g}\, \sigma^{ab} \partial_a \tilde \phi \partial_b \tilde \phi.
 \ee
 In this case, the contribution to the action \eqref{eq:actioncurved} from $\sigma^{ab} \tilde{R}_{ab}$ and the matter action $S_\mathrm{matt}$ are both separately compatible with index factorization and moreover exactly cancel each other. The individual index factorization of the two terms and the cancellation between $S_\mathrm{matt}$ and $\sigma^{ab} \tilde{R}_{ab}$ both stem from the fact that the matter Lagrangian for the free massless scalar is linear in the metric $g^{ab}$, which allows for the background value of the scalar action to be equal to $\sqrt{-g}\, g^{ab} \tilde{R}_{ab}$. In this case, the background value of the scalar becomes irrelevant to the gravity action, which in factorized form reduces to that given in \Eq{eq:actionfactorcurved}.

\subsection{Equations of Motion}

As we saw previously, the twofold Lorentz invariance of the action is directly manifest in the corresponding Feynman diagrams.   Moreover, this property should be exhibited by the equations of motion, i.e., the Einstein equations.  In this section, we compute the Einstein equations, to all orders in perturbation theory in our chosen field basis, about an arbitrary curved spacetime background. 
A priori, one can compute the equations of motion corresponding to field variations of $g_{ab}$, $\sigma_{ab}$, or $h_{ab}$, but all of these are related to each other by an appropriate Jacobian.

The Einstein equations with respect to $g_{ab}$ are of the standard form,
\be
 R_{ab} - \frac{1}{2} R\,g_{ab} = 8\pi G\,T_{ab}, \label{eq:EFE}
\ee
where we have defined the stress-energy tensor
\be 
T_{ab} = -\frac{2}{\sqrt{-g}} \frac{\delta (\sqrt{-g} \, {\cal L}_{\rm matt})}{\delta g^{ab}} 
\label{eq:Tab}
\ee
for matter Lagrangian ${\cal L}_\textrm{matt}$.  
Now, let us relate the usual Einstein equations to the equation of motion corresponding to the variation of $\sigma_{ab}$. The Jacobian relating $\sigma_{ab}$ and $g_{ab}$ is
\be 
J^{ab}_{\;\;\;\;cd}  = \sqrt{\frac{-g}{-\tilde{g}}} \frac{\delta g^{ab}}{\delta \sigma^{cd}} = \frac{1}{2}\left(\delta^a_c \delta^b_d + \delta^a_d \delta^b_c\right) - \frac{1}{D-2} g^{ab} g_{cd},\label{eq:Jacobian}
\ee
which has the structure of  the graviton propagator numerator in harmonic gauge. To obtain the equations of motion for $\sigma_{ab}$, we multiply \Eq{eq:EFE} by the Jacobian, yielding
\be 
 J^{cd}_{\;\;\;\;ab}\left(R_{cd} - \frac{1}{2}R\,g_{cd}\right) =  8\pi G\,J^{cd}_{\;\;\;\;ab} T_{cd}  \qquad \implies \qquad
  R_{ab} = 8\pi G\left(T_{ab} - \frac{1}{D-2} T g_{ab}\right),\label{eq:EFEcurved}
\ee
which are just another form of the Einstein equations.  Varying the action \eqref{eq:actioncurved} with respect to $\sigma^{ab}$, we obtain the equations of motion to all orders in perturbation theory,
\be
\begin{aligned}
 R_{ab} &= \frac{1}{2}\tilde\nabla_c\left(\sigma^{cd}\tilde{\nabla}_a\sigma_{bd} + \sigma^{cd}\tilde{\nabla}_b\sigma_{ad} - \sigma^{cd}\tilde{\nabla}_d\sigma_{ab}\right)  \\ &\qquad + \frac{1}{2}\left(\sigma^{ce}\sigma^{df} - \sigma^{cf}\sigma^{de}\right)\tilde{\nabla}_d \sigma_{ac}\tilde{\nabla}_f \sigma_{be}   +  \frac{1}{4} \tilde{\nabla}_a \sigma_{cd} \tilde{\nabla}_b \sigma^{cd} + \tilde{R}_{ab}.
 \label{eq:factorEeq}
\end{aligned}
\ee
A useful trick for handling the inverse matrix $\sigma^{ab}$ in the equations of motion is to introduce a constraint term $\lambda^a_b(\sigma_{ac}\sigma^{cb} - \delta_a^b)$, where $\lambda$ is a Lagrange multiplier. 
As consistency check, we can instead write $R_{ab}$ explicitly in terms of $g_{ab}$ for the perturbation expansion about flat spacetime, substituting in our field redefinition from \Eq{eq:pi}.  Indeed, in this case we obtain the same result as the flat-background limit of \Eq{eq:factorEeq}.

Let us momentarily consider the linearized Einstein equations in the flat-spacetime limit, 
\be
-\Box h_{ab} + \partial_a  \partial_c h^c_{\;\;b} + \partial_b \partial_c h^c_{\;\;a} = 16\pi G \left(T_{ab} - \frac{1}{D-2} T \eta_{ab}\right),\label{eq:linear}
\ee
where $\Box =  \eta^{ab}\partial_b \partial_a$. The left-hand side of \Eq{eq:linear} is nearly the same as the general, non-gauge-fixed linearized field equations in the so-called trace-reversed basis \cite{MTW}, but is missing the term $-\eta_{ab}\partial^c \partial^d h_{cd}$, which violates index factorization and which was removed by our gauge-fixing procedure. Note that we have not eliminated all of the available gauge freedom, since we can shift the coordinate functions $x^a$ by a perturbation $\delta x^a$ satisfying $\nabla_b \nabla_a \delta x^b = 0$. Equivalently, as noted in \Sec{sec:propagators}, we can send $h_{ab} \rightarrow h_{ab} + \partial_a \xi_b + \partial_b \xi_a$, as long as $\partial^a \xi_a = \text{constant}$, so that $\partial_a [h] = 0$ and the gauge condition \eqref{eq:GC} remains satisfied. In particular, for a radiative solution in which $T_{ab} = 0$, we can choose $\xi^a$ such that $\Box \xi^a = - \partial_b h^{ab}$, in which case we find that, after the shift, the perturbation satisfies $\partial^a h_{ab} = 0$. Hence, the vacuum equation reduces to the usual wave equation $\Box h_{ab} = 0$.

We now turn back to the general case of the Einstein equation for perturbation theory in an arbitrary background spacetime. To be as general as possible, we will for now ignore the issue of factorization and merely consider some implications of \Eq{eq:factorEeq}, the equation of motion for the gravity action \eqref{eq:actioncurved}. 
While \Eq{eq:actioncurved} appears with a tadpole in the graviton, $h^{ab}\tilde{R}_{ab}$, this is precisely canceled by additional tadpole terms generated by the matter action.  This is mandated by the equations of motion for the background, 
\be 
\tilde{R}_{ab} - \frac{1}{2}\tilde{R}\, \tilde{g}_{ab} = 8\pi G\,  \tilde{T}_{ab}, 
\ee
where $\tilde{R} = \tilde{R}_{ab}\tilde{g}^{ba}$. So as long as the background spacetime satisfies the Einstein equation, the tadpole in the action in \Eq{eq:actioncurved} is canceled.

For backgrounds with vanishing $\tilde{R}_{ab}$, we can also write the equations of motion associated with the action in \Eq{eq:actionfactorcurved} in terms of the fields $\Sigma_{AB}$ on the doubled spacetime, so that the factorization of indices in $h_{a \bar b}$ occurs automatically.  Doing so, we can rewrite \Eq{eq:factorEeq} as
\be
\begin{aligned}
[R_{AB}]_{\, x = \bar x }&= \bigg[\frac{1}{4}\tilde{\nabla}_C \left(\Sigma^{CD} \tilde{\nabla}_B \Sigma_{AD} + \Sigma^{DC}\tilde{\nabla}_A \Sigma_{DB} - \frac{1}{2}\Sigma^{CD}\tilde{\nabla}_D \Sigma_{AB} \right) \\ &\qquad + \frac{1}{8}\left(\Sigma^{EC}\Sigma^{FD} - \Sigma^{FC}\Sigma^{ED}\right) \tilde{\nabla}_D \Sigma_{AC} \tilde{\nabla}_F \Sigma_{EB} + \frac{1}{8}\tilde{\nabla}_A \Sigma_{CD} \tilde{\nabla}_B \Sigma^{CD}\bigg]_{\, x = \bar x} ,\label{eq:EFEfactorcurved}
\end{aligned}
\ee 
where $R_{AB}$ is the lift of the Ricci tensor into the $2D$-dimensional space.
We thus have a factorized form of the Einstein equations, valid to all orders in perturbation theory about an arbitrary curved vacuum background spacetime. Note, however, that if one simply varies the doubled-spacetime action in \Eq{eq:actionfactorcurved} with respect to $\Sigma_{AB}$, the resulting expression contains various errors in factors of two compared to the correct expression in \Eq{eq:EFEfactorcurved}, since the Lagrangian is integrated only over the diagonal spacetime $x = \bar x$. If one substitutes explicit expressions for $\Sigma_{AB}$ in terms of $h_{a \bar b}$ and then drops all bars, setting $x = \bar x$, then \Eq{eq:EFEfactorcurved} reduces to \Eq{eq:factorEeq} with $\tilde{R}_{ab} = 0$ as required.

\section{Conclusions}\label{sec:conclusions}

In this paper, we have described a systematic search for a pure gravity action exhibiting the twofold Lorentz symmetry suggested by the double copy relations. This property is manifested by two sets of indices, barred and unbarred, that are independently contracted and naturally parameterized by an auxiliary set of extra spacetime dimensions. By exploring the space of nonlinear field redefinitions and local gauge-fixing of the Einstein-Hilbert action discussed in \Sec{sec:FD}, we derived the twofold Lorentz invariant action described in \Sec{sec:factor}.   This action extends to an infinite family of actions related by twofold Lorentz invariant field redefinitions.  Some choices, e.g., the case of the Cayley-like basis explored in \Sec{sec:simplicity}, possess enhanced simplicity in terms of the reduced number of Lorentz invariant structures present in the action at each order in perturbation theory.

Because our results for flat-spacetime perturbation theory apply to all orders in the graviton field, they can be extended to curved spacetime.   In \Sec{sec:curved}, we derived a simple action for graviton perturbations around an arbitrary curved spacetime background in our field basis. Furthermore, we found that this action exhibited the same factorization properties for arbitrary Ricci-flat background spacetimes. We derived the Einstein equations in index-factorized form to all orders in the graviton about an arbitrary vacuum background and explored several interesting features they possess.

This work leaves a number of promising directions for future research.   First of all, while we introduced auxiliary spacetime dimensions simply as a convenient bookkeeping tool, it is likely that these can be derived from a truly extra-dimensional construction.  One path would be to understand how our action somehow arises as a truncation of double field theory that lifts the dilaton from the spectrum.  Alternatively, one could introduce dynamics governing fluctuations of the $D$-dimensional region within the doubled spacetime or smear out the delta function in \Eq{eq:actionfactor}, modifying the theory in the ultraviolet. 

Second, it would be illuminating to study the properties of graviton scattering amplitudes computed with this class of twofold Lorentz invariant actions.  Indeed, it has long been known that the properties of on-shell graviton scattering amplitudes enjoy improved high-momentum behavior from the study of BCFW recursion relations \cite{Britto:2004ap,Britto:2005fq} for general gauge and gravity theories  \cite{ArkaniHamed:2008yf,Cheung:2008dn}.  As discussed in \Ref{ArkaniHamed:2008yf}, these properties can be understand from a ``spin Lorentz symmetry'' that can be derived from the high-energy limit of these theories.  From this perspective, the results of this paper are a nonlinear generalization of this property beyond the high-energy limit.

Last but not least, a critical open question is whether and how our results relate directly to the double copy construction for scattering amplitudes in gauge theory and gravity.  
Here, it would be extraordinary to somehow reformulate our family of twofold Lorentz invariant gravity actions as two bona fide gauge theory copies.   The naive prescription---to simply substitute $h_{a \bar b} \sim A^{\vphantom{}}_a \bar{A}_{\bar b}$ at the level of Feynman vertices---is ambiguous since there are an infinite number of pure gravity actions from which one can start.  Nevertheless, we believe that a formulation likely exists, in part because the analogous puzzle has been understood for the double copy of effective field theories, where new representations of the nonlinear sigma model and special Galileon theories \cite{Cheung:2016prv} manifest these dualities as a symmetry of a cubic action.  In any case, this paper represents an initial step towards understanding the gauge and gravity double copy at the level of the action.

\begin{center} 
 {\bf Acknowledgments}
 \end{center}
 \noindent 
We thank Nima Arkani-Hamed, Zvi Bern, Sean Carroll, Roberto Percacci, Alan Weinstein, and Mark Wise for useful discussions and comments. C.C. is supported by a Sloan Research Fellowship and a DOE Early Career Award under Grant No. DE-SC0010255. G.N.R.~is supported by a Hertz Graduate Fellowship and a NSF Graduate Research Fellowship under Grant No.~DGE-1144469.

\appendix

\bibliographystyle{utphys}
\bibliography{factGR}

\end{document}